# The Impact of non-DLVO Forces on the Onset of Shear Thickening of Concentrated Electrically Stabilized Suspensions

Joachim Kaldasch* and Bernhard Senge

Technische Universität Berlin

Fakultät III: Lebensmittelrheologie

Königin-Luise-Strasse 22

14195 Berlin

Germany

Jozua Laven

Eindhoven University of Technology

Laboratory of Materials and Interface Chemistry

PO Box 513

5600 MB Eindhoven

The Netherlands

* author to who correspondence should be sent

**Abstract**

This paper exposes an extension of an activation model previously published by the authors. When particles arranged along the compression axis of a sheared suspension, they may overcome the electrostatic repulsion and form force chains associated with shear thickening. A percolation based consideration, allows an estimation of the impact of the force chains on a flowing suspension. It suggests that, similar to mode-coupling models, the suspension becomes unstable before the critical stress evaluated from the activation model is reached. The theory is applicable only to discontinuous shear thickening, and the predictions are compared with results from two experimental studies on aqueous suspensions of inorganic oxides; in one of them hydration repulsion and in the other hydrophobic attraction can be expected. It is shown that the incorporation of non-DLVO forces greatly improve predictions of the shear thickening instability.

# 1. Introduction

A number of theoretical attempts were made in the past to explain shear thickening in electrically stabilized colloidal suspensions. There are four main directions in the description of shear thickening.

One proposal for the source of shear thickening is that short-range lubrication forces are responsible for the formation of shear induced hydroclusters (Brady and Bossis (1985), Bender and Wagner (1989), Maranzano et al. (2001), Melrose (2003), Melrose and Ball (2004)). Another is that shear thickening is related to an order-disorder transition, where an ordered, layered structure becomes unstable above a critical shear rate (Hoffman (1974), Boersma et al. (1990),(1995), Hoffman (1998)).

Based on the idea of a stress-induced transition into a jammed state, a third approach describes the instability by taking advantage of a mode-coupling model, where the memory term takes into account the density, shear stress and shear rate (Cates et al. (1998), Holmes et al. (2003), (2005)).

Recently a fourth approach was proposed by Kaldasch et al. (2008) that is based on an activation model which suggests that particles in a sheared suspension may overcome the mutual electrostatic repulsion. As a result particle clusters in the flowing suspension are generated, leading to the shear thickening phenomenon. The interaction potential was described by the Derjaguin-Landau-Verwey-Overbeek (DLVO-) theory (Derjaguin and Landau (1941), Verwey and Overbeek (1948)), as the combined action of Van der Waals attraction and electrostatic repulsion.

However, with the advent of force measuring devices like the Surface Force Apparatus (SFA) and the Atomic Force Microscope (AFM), it has become possible to measure, with great sensitivity, the interaction forces between two surfaces down to molecular separations. As a result of these force measurements, serious limitations of the DLVO theory have come into light. Specifically, in aqueous solutions, depending on the situation, one has to take into account the extraneous short range attraction between hydrophobic surfaces, known as the hydrophobic force (Israelachvili and Pashley (1982), Rabinovich and Derjaguin (1988), Claesson and Christenson (1988)) and a repulsive hydration force appearing e.g. between silica surfaces (Peschel et al.(1982), Rabinovich and Derjaguin (1988), Grabbe, A. and Horn(1993)). The hydration force between silica surfaces might be attributed to the fact that hydrated cations bind onto the silica/water interface leaving a large volume of hydration layers on the surfaces of the silica particles (Song et al. (2005)). The origin of the hydrophobic forces is unclear and there is a debate whether the deviation from continuum behavior should be attributed to molecular properties of the liquid or to surface interactions. One explanation is based on the effect the particle/water interface may have on the structure of the water in close proximity to that interface (Rabinovich and Yoon (1994)). This local deviation from the order in the bulk decays with increasing distance from the interface and gives rise to a short-range interaction with the same decay length as the order profile. Another proposal uses classical continuum electrostatics to illustrate how electrostatic correlations, arising from a surface-induced perturbation in the fluid next to a hydrophobic surface, could give rise to a long-range force ( Podgornik (1989)). A more qualitative suggestion is that the formation of vapor cavities between the hydrophobic surfaces plays some role in generating the force (Claesson and Christenson (1988)). None of these theories accounts for all of the experimental results. Although investigators disagree over the origin of the hydrophobic force, its existence has been confirmed independently by various research groups (Rabinovich et al. (1982), Claesson and Christenson (1988)).

    The main goal of this paper is to extent the rheological model of shear thickening supposed by Kaldasch et al.(2008, 2009) in two respects. First, non-DLVO forces are taken into account. Second, we want to study how density perturbations (hydro-clusters) near the shear thickening instability change the rheological response of the sheared suspension. A

comparison of the model predictions with experimental investigations on aqueous alumina- and silica suspensions shows the applicability of the model.

## 2. The Model

We want to consider a concentrated, electrically stabilized, colloidal suspension of monodisperse spherical particles. The DLVO-theory states that the total two-particle interaction can be expressed as a sum of the electrical double-layer interaction $U_{el}$ and the Van der Waals attraction $U_{vdW}$. However, experimental investigations showed the existence of forces not considered in the DLVO-theory, known as hydrophobic and the hydration forces. In order to include these contributions, the DLVO-theory may be extended by writing the total interaction potential as a function of the surface-to-surface distance $h$:

$$U(h) = U_{el}(h) + U_{vdW}(h) + U_S(h)$$

(1)

where the structural potential $U_S$ represents either a hydrophobic or hydration potential.

The electrostatic repulsion energy of the particles is due to the overlap of the electric double layer. In numerous cases, a simple equation derived by Hogg, Healy and Fuerstenau (1966) was found to be a suitable approximation, which for our case can be expressed for constant surface potential as a function of the surface-to-surface distance $h$ by:

$$U_{CP}(h) = 2\pi \varepsilon_0 \varepsilon_r a \Psi_0^2 \ln(1 + e^{-\kappa h})$$

(2)

and for a constant surface charge approach of the particles as:

$$U_{CC}(h) = -2\pi \varepsilon_0 \varepsilon_r a \Psi_0^2 \ln(1 - e^{-\kappa h})$$

(3)

where we approximate $\Psi_0$, the surface potential at infinite interparticle distance, by the $\zeta$-potential. The parameters $\varepsilon_0$ and $\varepsilon_r$ are the absolute and relative dielectric constants and $a$ is the particle radius. The Debye reciprocal length $\kappa$ is defined by:

$$\kappa = \sqrt{\frac{2 C_S N_A Z^2 e_0^2}{\varepsilon_0 \varepsilon_r k_B T}}$$

(4)

where $e_0$ is the elementary electric charge, $N_A$ the Avogadro- number, $Z$ the ionic charge number and $C_S$ is the salt concentration.

The non-retarded Van der Waals attraction between two spheres can be taken into account by:

$$U_{vdW}(h) = -\frac{Aa}{12h}$$

(5)

where $A$ is the effective Hamaker constant determined by the dielectric constants of the solvent-particle combination.

The short-range surface forces observed at separation distances less than *10 nm* can be described by an empirical single exponential law (Israelachvili and Pashley (1982)):

$$U_S(h) = \pm a C_0 D_0 \exp\left(\frac{-h}{D_0}\right)$$

(6)

where $C_0$ is a pre-exponential factor and $D_0$ is called the decay length. The minus sign refers to hydrophobic attraction and the plus sign to hydration repulsion.

**The critical stress**

In a suspension disturbed by simple shear, with a shear stress $\sigma$, a pair of colloidal particles may overcome the mutual repulsion of an energy barrier, $U_B$, along the compression axis of the flow. In a meanfield model this activated process takes place with a frequency

$$f = f_0 p$$

(7)

where *p* is the transition probability for the formation of a particle pair:

$$p = \exp\left(-\frac{U_B - \sigma V^*}{k_B T}\right)$$

(8)

while $k_B$ is the Boltzmann constant, $T$ the temperature, $\sigma$ the applied stress and $V^*$ the activation volume. The parameter $f_0$ represents the average collision rate of particles along the compression axis in the absence of flow. $f$ has its maximum at $p_C=1$.

The activation barrier is formed by the two-particle interaction potential:

$$U_B = U(h_{max}) - U(h_0)$$

(9)

where $h_0$ is related to the equilibrium particle distance and the maximum of the interaction potential, at $h_{max}$, can be obtained by

$$\frac{\partial U(h)}{\partial h} = 0; \frac{\partial^2 U(h)}{\partial h^2} < 0$$

(10)

For randomly distributed particles the preferred surface to surface distance, which is identical to $h_0$, is governed by the volume fraction $\Phi$:

$$h_0(\Phi) \cong 2a\left(\left[\frac{\Phi_m}{\Phi}\right]^{1/3} - 1\right)$$

(11)

where $\Phi_m=0.64$ is the maximum packing density of a hard sphere suspension.

$V^*$ represents the volume associated with this activation process, which is of the order of the free volume per particle:

$$V^* = \frac{4}{3}\pi a^3 \frac{\Phi_m}{\Phi}$$

(12)

We want to follow Cates et al. (1998) and assume that shear thickening is associated with a jamming transition. The jamming transition is characterized by the occurrence of force chains.

The key idea of this model is the assumption, that a force chain corresponds to a chain of bonded particles arranged along the compression axis of the sheared suspension. The chance that a bond is present between two particles along the compression axis is given by Eqs.(7) and (8). An infinite chain occurs only, if the chance for a bond becomes of the order $p \approx p_C = 1$. From Eq.(8) follows for the critical stress in continuous shear thickening (Kaldasch et al. (2008)):

$$\sigma_C = \frac{U_B}{\frac{4}{3}\pi a^3 \frac{\Phi_m}{\Phi}}$$

(13)

The critical stress $\sigma_C$ can be evaluated for a constant potential (CP) and a constant stress (CC) approach between the colloidal particles. According to Kaldasch et al. (2008) a critical stress, which we want to denote as transitional stress, $\sigma_t$, determines whether Eq. (2) of the CP case (if $\sigma_C < \sigma_t$) or Eq. (3) of the CC case ($\sigma_C > \sigma_t$) should be used when evaluating

$U_B$ in Eq.(13) for two particles that have CP interaction under static conditions. Its value is given by:

$$\sigma_t = \frac{6\eta_S D}{4a^2}$$

(14)

where $D$ is the diffusion constant of the ions given by

$$D = \frac{k_B T}{6\pi\eta_S a_i}$$

(15)

$a_i$ is the radius of the ions and $\eta_S$ is the viscosity of the solvent medium.

**The percolation approach**

We want to confine our interest to the generation of flow-induced clusters at stresses close to $\sigma_C$, where the growth process of hydroclusters along the compression axis takes place much faster than their rotation and break-up. The probability for bonds between the colloidal particles along the compression axis can be written as:

$$p = \exp\left(-\frac{V^*}{k_B T}(\sigma_C - \sigma)\right)$$

(16)

The formation of the bonds is a dynamic process. However, let us take a picture of the sheared suspension at different stresses and determine the typical linear cluster size along the compression axis, known as correlation length ξ. Since the bond-formation probability given by Eq.(16) increases with increasing stresses, it can be expected that ξ of the (hydro-)clusters becomes very large close to σ$_C$. The key idea of this approach is to consider these elongated clusters as force chains and treat them as one-dimensional objects. In this case the random formation of bonds between the colloidal particles along the compression axis can be treated as a bond-percolation process. In one dimension this problem can be solved exactly (Stauffer(2003)). The percolation approach suggests that at $p=p_C$, an infinite connected cluster should be present, while for $p<p_C$ only finite force chains occur. This approach implies that the suspension volume fraction must be high enough to form infinite clusters. Expanding Eq.(16) near the transition, we obtain:

$$p_C - p = A(\sigma_C - \sigma)$$

(17)

with

$$A = \frac{V^*}{k_B T}$$

(18)

The density correlation function along the compression axis as a function of the distance $x$ reads (Stauffer(2003)):

$$g(x) = p^x$$

(19)

which for p<1, turns into

$$g(x) = e^{-\frac{x}{\xi}}$$

(20)

with the correlation length

$$\xi = -\frac{1}{\ln(p)} \cong \frac{1}{p_C - p} = \frac{1}{A(\sigma_C - \sigma)}$$

(21)

close to $p_C$. The correlation length of the hydro-clusters along the compression axis diverges on approaching the critical stress. This divergence leads, however, to a considerable decrease of the dynamics of the sheared suspension near the instability, known as critical slowing down. The conventional theory of critical slowing down suggests that the relaxation time of the clusters scale as:

$$\tau = \lambda \xi^z$$

(22)

while $z$ is a dynamical critical exponent and $\lambda$ a free parameter.

**The mechanical properties**

Colloidal suspensions usually exhibit shear thinning for small shear rates $\dot{\gamma}$. We want to describe a sheared suspension near the shear thickening transition as a visco-elastic continuous medium. The visco-elastic response is modeled as:

$$\sigma = K_1 \dot{\gamma}^n + K_2 \gamma^n$$

(23)

where $\gamma$ is the deformation, $n$ the power law exponent of the shear thinning suspension and $K_1, K_2$ are free parameters. This equation corresponds to the Kelvin–model with a single relaxation time, where $K_1 = \eta$ is the apparent viscosity and $K_2 = G$ the elastic modulus of the unperturbed suspension, if $n=1$. For $\sigma=0$, a deformation relaxes according to:

$$\gamma \sim \exp\left(-\left(\frac{K_2}{K_1}\right)^{1/n} t\right)$$

(24)

where the internal relaxation time corresponds to

$$\tau^n = \frac{K_1}{K_2}$$

(25)

With this result Eq.(23) can be rewritten for an applied constant stress:

$$\dot{\gamma}^n = \frac{1}{\tau^n}\left(\frac{\sigma}{K_2} - \gamma^n\right)$$

(26)

As discussed above, approaching the jamming transition, the critical slowing down of the clusters let the internal relaxation time diverge. Due to this over-damped dynamics close to the shear thickening transition, elastic deformations can be neglected, $\gamma^n << \sigma/K_2$. Since the relaxation time close to the jamming transition must be given by Eq. (22), we obtain for the stress dependent shear rate:

$$\dot{\gamma} = Q(\sigma_C - \sigma)^z \sigma^q$$

(27)

where

$$Q = \frac{A^z}{\lambda K_2^q}$$

(28)

and $q=1/n$. From Eq.(27) follows that on approaching $\sigma_C$, the shear rate exhibits a maximum at the stress

$$\sigma_m = \frac{q\sigma_C}{z+q}$$

(29)

and the corresponding shear rate:

$$\dot{\gamma}_m = Qq^q z^z \left(\frac{\sigma_C}{z+q}\right)^{z+q}$$

(30)

In qualitative agreement with mode-coupling models the present theory predicts that the shear rate decreases for $\sigma > \sigma_m$ (Holmes et al. (2003)). In this stress range the suspension is not mechanically stable, since $d\sigma/d\dot{\gamma} < 0$ and the present model is not applicable. Thus the theory predicts a discontinuity between a stable low shear stress branch for $\sigma \leq \sigma_m$, and an unstable branch for $\sigma > \sigma_m$. Hence a discontinuous shear thickening occurs at $\sigma_m$. Experimental investigations indicate that the suspension is governed by a stick-slip-motion within the unstable regime (Boersma et al. (1990)). The present model suggests that the elastic properties of the force chains dominate the rheological response of the suspension for $\sigma > \sigma_C$.

## 4. Comparison with Experimental Investigations

We want to compare our theoretical predictions with results of two experimental studies of discontinuous shear thickening, in which hydration and hydrophobic forces can be studied.

Hydrophobic forces can be expected in aqueous alumina suspensions investigated by Zhou et. al. 2001. The aqueous alumina suspensions AKP-15L were studied at constant volume fraction $\Phi=0.56$ with a salt concentration 0.01 M $KNO_3$ and different pH-values. The characteristic properties of the samples are summarized in Table 1. For the calculation of the attractive contribution between the particles, the Hamaker constant was taken to be $A=4.1*10^{-20}$ *J*, while the hydrophobic parameters: $C_0=-30$ *mJ/m²* and $D_0=1.2$ *nm* were obtained from what Israelachvili and Pashley (1982) found for hydrophobic particles. The interaction potential at constant charge is displayed in Fig. 1, for *pH=5.5*. Since the maximum of the potential is very close to the surface, surface forces have a considerable impact on its magnitude and therefore also on the onset of shear thickening as suggested already by Franks et al. (2000).

Note that the present model predicts that the critical stress of an electrically stabilized suspension is given by Eq.(13). However, the percolation approach suggests that the suspension shows a discontinuous shear thickening not the critical stress, but the maximum stress given by Eq.(29). This equation contains a power law exponent $q$ and the critical exponent $z$ as unknown parameters, not derived in this simple theory. Therefore both parameters have to be specified form experimental data, before an experimental critical stress can be evaluated.

In order to determine these unknown parameters, we want to consider the experimentally obtained shear rates for a series of increasing shear stresses of the sample AKP-15L as displayed in Fig.2. Since critical slowing down occurs only close to the transition, $q$ can be specified at small shear stresses in the shear thinning rage of the flow curve, far from the transition, when elastic contributions to the viscosity can be neglected. From Eq.(27) follows that for $\sigma \ll \sigma_C$, the shear rate can be approximated by $\dot{\gamma} \sim \sigma^q$, which allows an estimation of the power law exponent. From Fig. 2 for the sample AKP-15L a value $q=1.9 \pm 0.02$ (dotted line) was found.

Once $q$ has been determined, the critical exponent, $z$, can be specified by using two points of the theoretical flow curve, as defined by Eq.(27). One point is the maximum stress $\sigma_m=45\ Pa$ at shear rate $\dot{\gamma}_m=300s^{-1}$, and the second is an arbitrary data point at small shear stresses. Substituting these two points in Eq.(27) and Eq.(29), the critical exponent can be obtained from

$$\frac{\dot{\gamma}_m}{\dot{\gamma}_0}\left(\frac{\sigma_0}{\sigma_m}\right)^q = \left(\frac{\sigma_C - \sigma_m}{\sigma_C - \sigma_0}\right)^z = \left(\frac{z}{z + q\left(1 - \frac{\sigma_0}{\sigma_m}\right)}\right)^z$$

**(31)**

For the third data point $\sigma_o=0.6\ Pa$ and $\dot{\gamma}_0=0.2s^{-1}$ a numerical evaluation gives $z=0.77$. The average over the next 10 data points of Fig.2 is $z=0.97 \pm 0.13$ with a corresponding experimental critical stress $\sigma_C=68 \pm 3\ Pa$ (Eq.(29)). For these parameters Eq.(27) with $\sigma=\sigma_m$ and $\dot{\gamma}=\dot{\gamma}_m$ suggests that $Q=0.009 \pm 0.001$.

The theoretical critical stress at constant charge including surface forces evaluated from Eq.(13) is $\sigma_C=70.5\ Pa$, which is in good agreement with the experimental result. From equilibrium thermodynamics, it can be expected that density fluctuations are independent of the particle properties. On that condition a good approximation of the maximum stress can be given by:

$$\sigma_m = \frac{q}{1+q}\sigma_C$$

(32)

which we want to apply here independent of the studied system. The corresponding flow curve, Eq.(27), is displayed in Fig.2 for $q=1.9$.

At constant volume fraction of the samples, $q$ is considered also constant, and the maximum critical stress, $\sigma_m \approx 0.66\, \sigma_C$, where $\sigma_C$ is given by Eq.(13) at constant charge. The model predictions of the aqueous alumina suspensions investigated by Zhou et al. (2001) are in good agreement with the experimental data as displayed in Fig. 3, while for comparison the dotted line indicates the maximum stress, $\sigma_m$, at constant charge based only on DLVO forces (Kaldasch et al. (2008)).

We want to consider another experimental study performed by Franks at. al. (2000) on aqueous silica suspensions. In this system hydration repulsion can be expected (Song et al. (2005)). The silica particles involved are nearly monodisperse, nearly spherical with a diameter of 1 μm. The samples were studied at a constant volume fraction $\Phi=0.58$ and salt concentration *0.01 M NaCl*, at different *pH* values. The suspensions exhibited shear thickening at stresses summarized in Table 2. Empirically, the decay length of the hydration repulsion is usually of the order: $D_0 \approx 0.6$-$1.1$ nm for 1:1 electrolytes, while $C_0$ depends on the hydration of the surfaces and is $C_0 \approx 3$-$30$ mJ/m$^2$ (Paunov et al. (1996)). The power law exponent of the shear thinning part of the flow curve was nearly constant, $q=1.3$ (Franks (2008)). The fat line in Fig 4 obtained from Eq.(32) represents a fit to the experimentally obtained data with the parameter $C_0=30$ mJ/m$^2$ and $D_0=0.9$nm of the hydration repulsion. Note that the fit parameters are in the expected parameter range.

## 3. Conclusions

The critical stress, $\sigma_C$, as predicted by the model corresponds to the stress necessary to overcome the mutual repulsion between two particles and form an infinite force chain. However force chains are unstable against shear perturbations, since their relaxation time diverges on approaching the critical stress. The relaxation time can be estimated from a percolation approach, if we assume that the hydroclusters are highly elongated along the compression axis. Similar to mode-coupling models, suggesting an S-shaped curve of the stress-shear rate dependence, the present model predicts that the suspension becomes already unstable at a maximum stress, $\sigma_m < \sigma_C$. The extension of the DLVO-activation model by taking into account structural forces improves the prediction of the shear thickening instability considerably, as was shown by a comparison with experimental investigations on aqueous suspensions with significant hydrophobic and hydration forces. Note that the present model is applicable only to discontinuous shear thickening.

A problem of the this theory is that after having overcome the mutual repulsion, colloidal particles should remain trapped in the primary minimum of the interaction potential. We want to emphasize, that the surface-to-surface distances, $h_{max}$, at the maximum of the continuum description of the interaction potential, is very close to the surface. As summarized in Table 1 and Table 2, $h_{max}$ is usually a few hundred pico-meter. It becomes of the order of the size of the largest atoms of the particles involved, which is *~200 pm* in diameter for a Si-atom and *~250 pm* for an Al-atom. This result leads us to speculate that the reversibility of shear thickening may be due to the surface roughness of the colloidal particles. The model suggests that orthokinetic coagulation (Smoluchowski (1912), Friedlander (2000)) and shear thickening belong to the same effect**.** Once the particles have overcome the mutual repulsion they are in a bounded state, and the difference between reversible shear thickening and irreversible shear aggregation is the magnitude of the potential barrier $U_{Bound}$ of the bounded particle to become unbounded. In the case of shear thickening, $h_{max}$ is of the order of the atomic constituents and the surface roughness prohibits an approach far beyond this distance. It can be expected that with suspensions that exhibit shear thickening $U_{Bound} < k_B T$, because after ceasing the shear perturbation a shear thickening suspension relaxes into a stable suspension due to thermal activation. For the case of a shear aggregated suspension, $U_{Bound} >> k_B T$ and the suspension remains in the coagulated state.

Inevitably, the model requires that the two-particle interaction potential has a maximum at a finite distance. Unlike electrostatically interacting particles, particles stabilized by adsorbing polymer exhibit a potential, which consists only of a shallow attractive minimum followed by a rapid increase as the surfaces of the colloidal particles approach each other closely. Shear thickening in suspensions stabilized by absorbed polymers, must be due to an alternative mechanism as discussed in Maranzano et al. (2001) and Krishnamurthy et al. (2005) .

Note that the present model is in agreement with the assumption of a shear-induced jamming transition as suggested by Cates et al. (1998), which implies that a periodic structure must suffer from a structural transition, as purposed as an explanation for shear thickening by Hoffman (1974). It can be expected that the understanding of the shear thickening instability can be improved substantially when the rheological characterization of the critical stresses is supplemented with a more precise determination of the interaction potential of the colloidal particles by direct force measurements. Therefore more experimental investigations with accurately characterized model systems are necessary.

**Tables**

| Suspension | pH | $\zeta$ [mV] | $h_{max}$ [pm] | $\sigma_C$ [Pa] | $\sigma_m$ [Pa] (theory) | $\sigma_m$ [Pa] (experiment) |
|---|---|---|---|---|---|---|
| AKP 15L | 5.0 | 96 | 116 | 76 | 49.8 | 50 |
| AKP 15L | 5.5 | 90 | 136 | 58 | 37.7 | 38 |
| AKP 15L | 6.0 | 80 | 188 | 41 | 20.4 | 30 |
| AKP 15L | 6.4 | 73 | 258 | 22 | 10.7 | 12 |
| AKP 15L | 6.8 | 64 | 2191 | 8 | 3.4 | 6 |

**Table 1.** Characteristic data of the alumina suspensions AKP 15L investigated by Zhou et al. (2001). *T=298 K, a=0.425 µm, C=10 mol/m³, $\varepsilon_r$=80.37, A=4.1*10$^{-20}$ J, Z=1, q=1.9, z=1.*

- 22 -| Suspension | pH | $\zeta$ [mV] | $h_{max}$ [pm] | $\sigma_C$ [Pa] | $\sigma_m$ [Pa] (theory) | $\sigma_m$ [Pa] (experiment) |
|---|---|---|---|---|---|---|
| silica in water | 8.35 | 60 | 40 | 68 | 39 | 34 |
| silica in water | 6.85 | 44 | 66 | 41 | 24 | 24 |
| silica in water | 5.50 | 30 | 102 | 26 | 15 | 18 |
| silica in water | 4.50 | 13 | 150 | 18 | 10 | 9 |
| silica in water | 1.90 | 0 | N/A | N/A | N/A | N/A |

**Table 2.** Characteristic data of the silica suspensions investigated by Franks et al. (2000). $T=298$ K, $a=0.5$ μm, $C=10$ mol/m$^3$, $\varepsilon_r=80.37$, $A=8.3*10^{-21}$ J, $Z=1$, $q=1.3$, $z=1$.

**Figures**

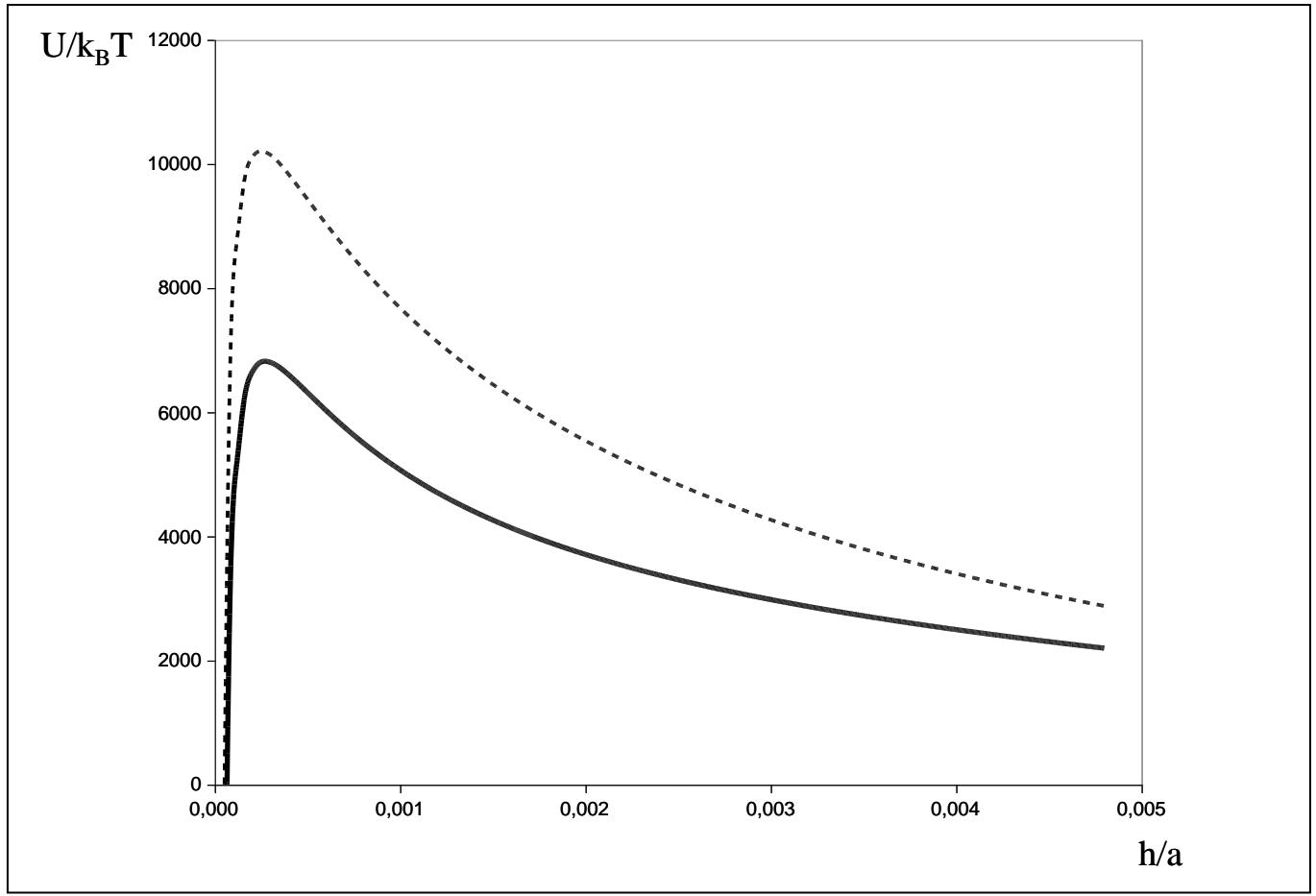

**Figure 1.** Scaled interaction potential at constant charge as a function of the two-particle surface-to-surface distance *h/a* of sample AKP-15L (*a=425 nm*) with the flow curve displayed in Fig.2. While the solid line represents the total potential, the dotted line indicates the interaction potential without hydrophobic surface forces.

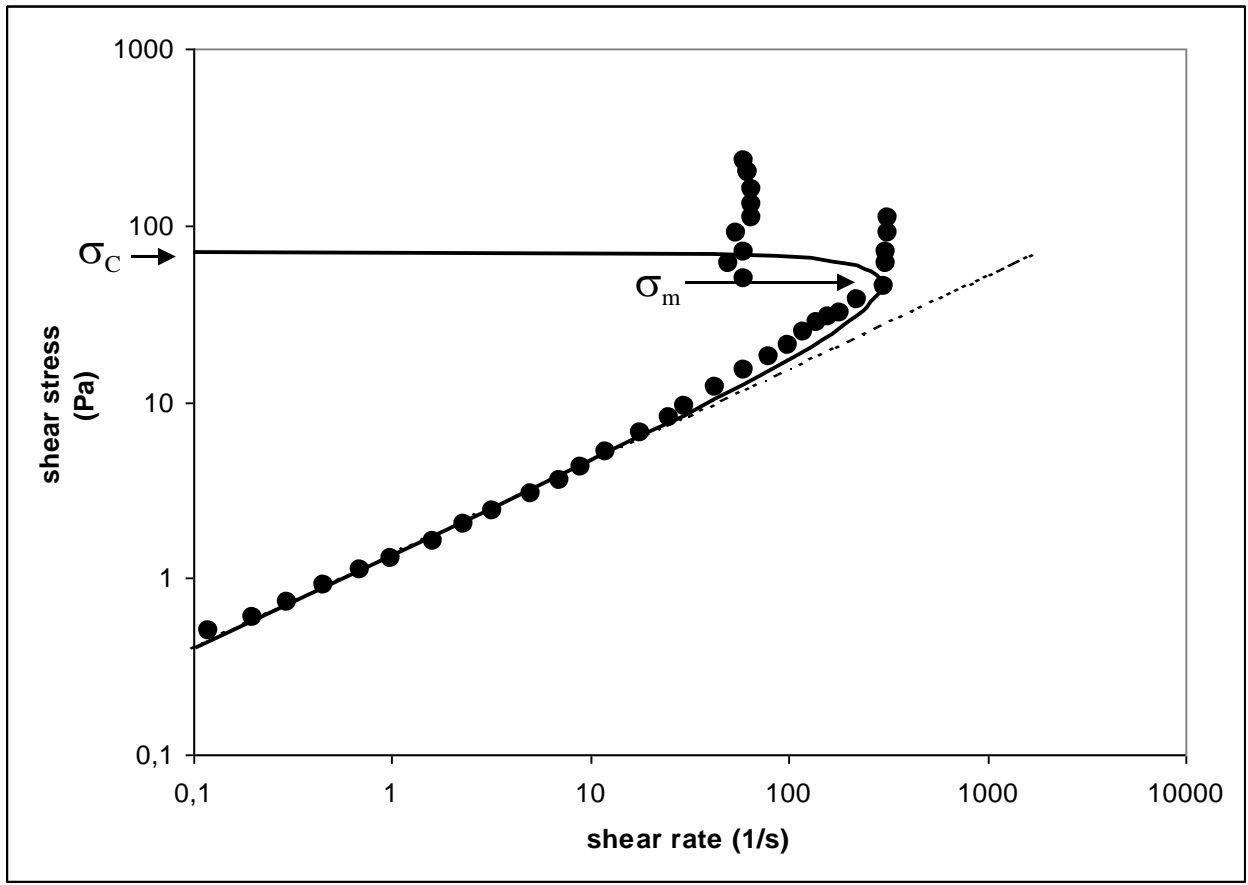

**Figure 2.** Shear stress as a function of the shear rate of the alumina suspension AKP-15L $\Phi=0.56$ at *0.005 M* KNO$_3$ and *pH=5.5*, investigated by Zhou et al. (2001) (circles). The dotted line indicates that the sample exhibits shear thinning with *q=1.9* for small stresses. The fat line corresponds to Eq.(27) with *Q=0.008* and *z=1*. The suspension can be expected to be unstable for stresses, $\sigma > \sigma_m$, while $\sigma_C$ is the critical stress.

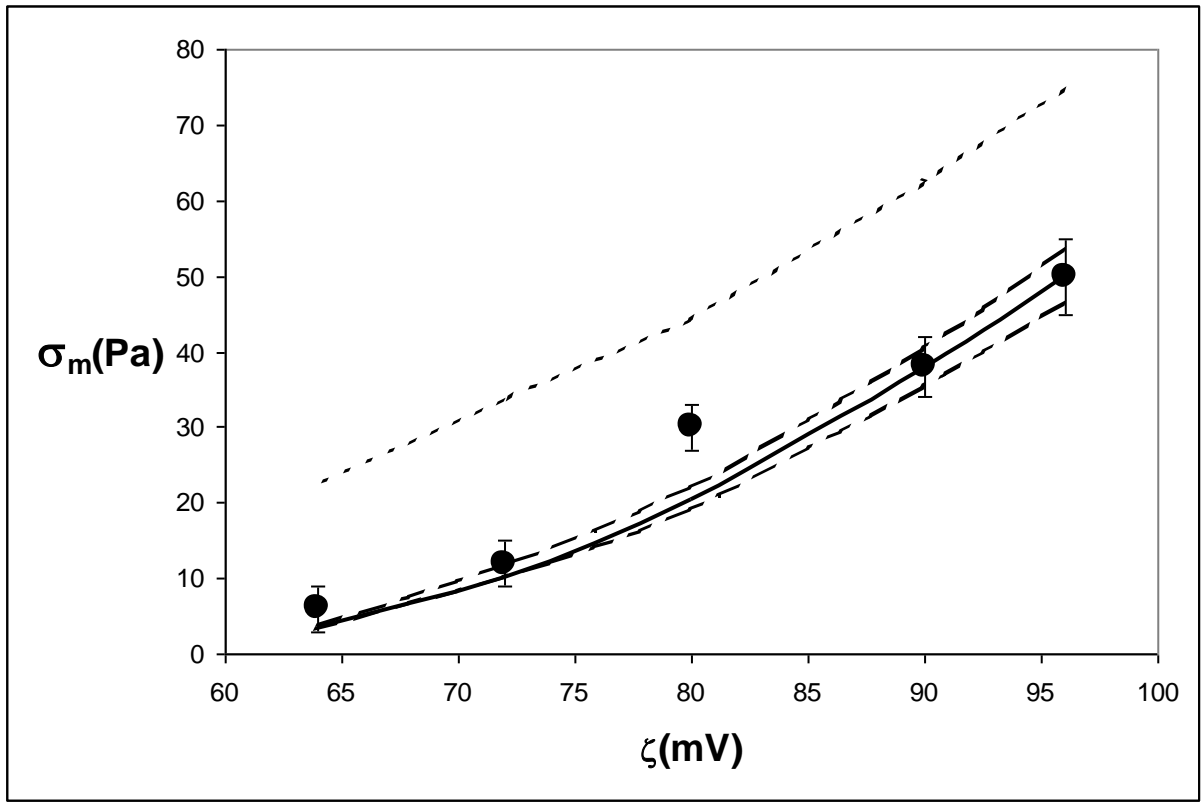

**Figure 3.** The maximum shear stress, σ$_m$, at constant charge taking into account the hydrophobic force (fat line) as a function of the surface potential (ζ-potential) of the alumina suspension AKP-15L investigated by Zhou et al. (2001) (circles). The dotted line indicates the outcome of the model without surface forces. The dashed lines show if *z* varies *z=1±0.2*.

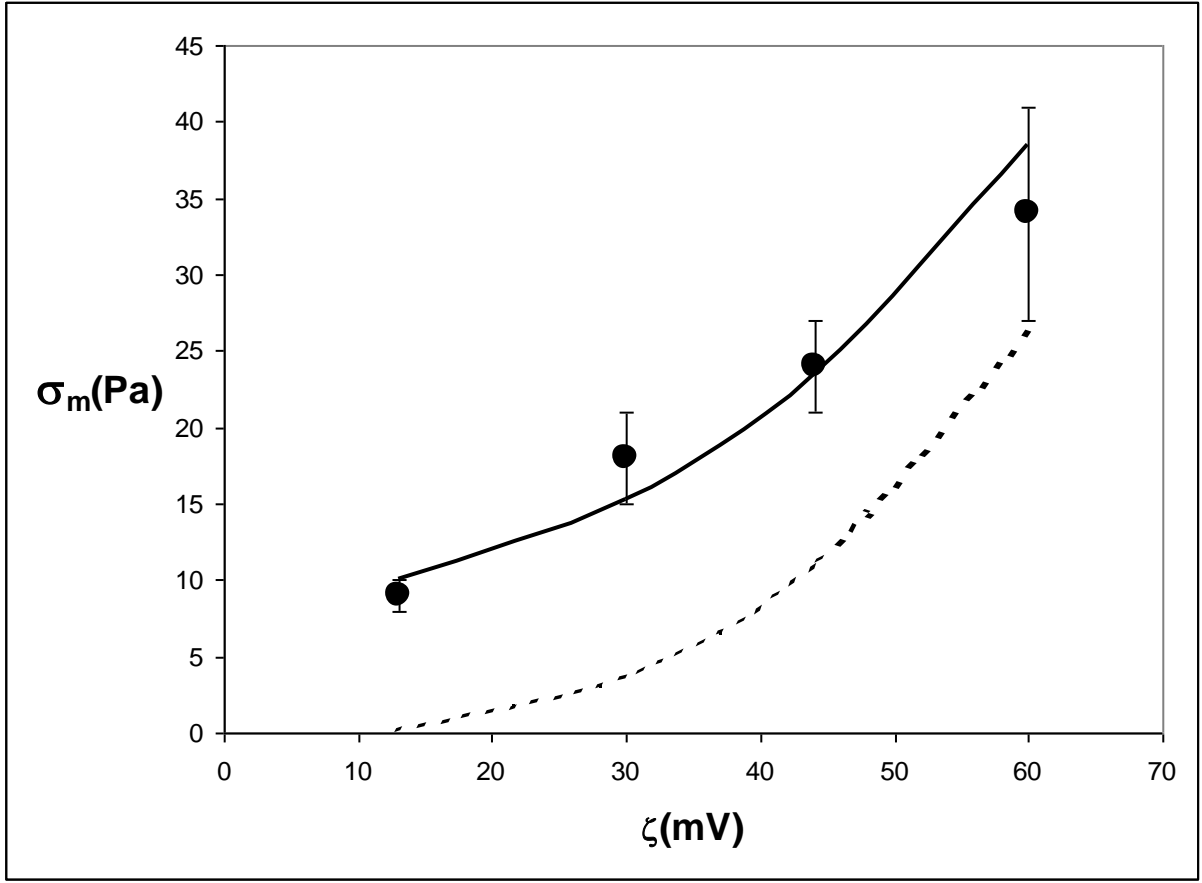

**Figure 4.** The maximum shear stress, $\sigma_m$, at constant charge taking into account the hydration force (fat line) as a function of the surface potential ($\zeta$-potential) of the silica suspensions investigated by Franks et al. (2000) (circles). The dotted line indicates the outcome of the model without surface forces.

**Literature**

Boersma WH, J Laven, HN Stein (1990) Shear Thickening (Dilatancy) in Concentrated Dispersions. AIChE J. 36: 321-332

Bender JW, NJ Wagner (1989) Reversible shear thickening in monodisperse and bidisperse colloidal dispersions. JU. Rheol. 33:329-366

Brady JF, G Bossis (1985) The rheology of concentrated suspensions of spheres in simple shear flow by numerical simulation. J. Fluid Mech. 155:105-129

Cates ME, JP Wittmer, J-P Bouchaud, P Claudin (1998) Jamming, Force Chains, and Fragile Matter. Phys. Rev. Lett. 81: 1841−1844

Claesson, PM, HK Christenson(1988) Very long range attractive forces between uncharged hydrocarbon and fluorocarbon surfaces in water. J. Phys. Chem. 92: 1650-1655

Derjaguin BV, LD Landau (1941) Theory of stability of highly charged lyophobic sols and adhesion of highly charged particles in solutions of electrolytes. Acta Physicochim. URSS. 14: 633-652

Friedlander SK (2000) Smoke, Dust, and Haze. Fundamentals of Aerosol Dynamics. Oxford University Press, New York.

- 28 -
Franks VG, Z Zhou, NJ Duin, DV Boger (2000) Effect of interparticle forces on shear thickening of oxide suspensions. J. Rheol. 44: 759-779

Grabbe A, RG Horn (1993) Relationship between interfacial forces measured by colloid-probe atomic force microscopy and protein resistance of poly(ethylene glycol)-grafted poly(Llysine) adlayers on metal oxide surfaces. J. Colloid Interface Sci. 157: 375-383

Hoffman RL (1974) Discontinuous and Dilatant Viscosity Behavior in Concentrated Suspensions. II. Theory and Experimental Tests. J. Coll. Interface Sci. 46: 491-506

Hoffman RL (1998) Explanations for the cause of shear thickening in concentrated colloidal suspensions. J. Rheol. 42: 111-123

Hogg R, TW Healy, DW Fuerstenau (1966) Mutual coagulation of colloidal dispersions. Trans. Faraday Soc. 62: 1638 –1651

Holmes CB, M Fuchs, ME Cates (2003) Jamming transitions in a mode-coupling model of suspension rheology. Europhys Lett. 63: 240-246

Holmes CB, ME Cates, M Fuchs, P Sollich (2005) Glass transitions and shear thickening suspension rheology. J. Rheol. 49: 237-269

Israelachvili J, RM Pashley (1982) The hydrophobic interaction is long range, decaying exponentially with distance. Nature 300: 341-342

Kaldasch J, B Senge, J Laven (2008) Shear Thickening in electrically stabilized colloidal Suspensions. Rheol. Acta 47: 319-323

Kaldasch J, B Senge, J Laven (2009) Shear thickening in electrically stabilized non-aqueous colloidal suspensions. 19(2):

Maranzano BJ, NJ Wagner (2001) The effects of particle size on reversible shear thickening of concentrated dispersions. J. Chem. Phys. 23(114):10514-10527

Melrose JR (2003) Colloid flow during thickening – a particle level understanding for core-shell particles. Faraday Discuss 123: 355-368

Melrose JR, RC Ball (2004) Contact Networks in Continuously Shear Thickening Colloids. J. Rheol. 48(5): 961-978

Otsubo Y, M Horigome (2003) Effect of surfactant adsorption on the rheology of suspensions flocculated by associated polymers. Korea-Asutralia Rheol. J. 15(4): 179-185

Paunov VN, RI Dimova, PA Kralchevsky, G Broze, A Mehreteab (1996) The hydration repulsion between charged surfaces as an interplay of volume exclusion and dielectric saturation effects. J. Colloid & Interface Science 182**:** 239-248

Peschel G, P Belouschek, MM Muller, RM Muller, R Konig (1982) The interaction of solid surfaces in aqueous systems. Colloid & Polymer Science 260(4): 444-451

Podgornik R (1989) Electrostatic correlation forces between surface specific ionic interactions. J. Chem. Phys. 91: 5840-5849

Rabinovich YI, BV Derjaguin, NV Churaev (1982) Direct measurements of long-range surface forces in gas and liquid media. Adv. Colloid Interface Sci. 16: 63-78

Rabinovich YI, BV Derjaguin (1988) Interaction of hydrophobized filaments in aqueous electrolyte solutions. Colloids and Surfaces 30: 243-251

Rabinovich YI, R-H Yoon (1994) Use of Atomic Force Microscope for the Measurements of Hydrophobic Forces between Silanated Silica Plate and Glass Sphere. Langmuir 10: 1903-1909

Smoluchowski M (1917) Versuch einer mathematischen. Theorie der Koagulationskinetic kolloider Lösungen. Z. Phys. Chem. 92: 129-168

Stauffer D, A Aharony (2003) Introduction to Percolation Theory. Taylor & Francis Group, UK

Song S, C Peng, MA Gonzalez-Olivares, A Lopez-Valdivieso, T Fort (2005) Study on hydration layers near nanoscale silica dispersed in aqueous solutions through viscosity measurement . J. Colloid Interface Sci. 287: 114-120

Verwey EJW, JTG Overbeek (1948) Theory of the Stability of Lyophobic Colloids. Elsevier, Amsterdam

Zhou Z, PJ Scales, DV Boger (2001) Chemical and physical control of the rheology of concentrated metal oxide suspensions. Chem. Engineering Sci. 56: 2901-2920